\documentclass[intlimits,twoside,a4paper]{article}
\usepackage{subfigure}
\usepackage{amsmath,amssymb}
\usepackage{graphicx}

\usepackage[T2A]{fontenc}
\usepackage[cp1251]{inputenc}

\usepackage[eqsecnum]{cmpj2}

\issue{2013}{16}{3}{33802}
\doinumber{10.5488/CMP.16.33802}

\title[Strain energy calculations of hexagonal boron nanotubes]%
{Strain energy calculations of hexagonal boron nanotubes: An ab-initio approach}%
\author[S.K. Jain, P. Srivastava]{S.K. Jain\thanks{E-mail: sandeepjain@iiitm.ac.in, phone: +91-751-2449814, Fax: +91-751-2449814}\ , P. Srivastava} 
\address{Nanomaterials Research Group, Computational Nanoscience and Technology Lab (CNTL), \\ %
ABV-Indian Institute of Information Technology and  Management, Gwalior--474015, India}%

\date{Received September 3, 2012, in final form February 21, 2013}
\authorcopyright{S.K. Jain, P. Srivastava, 2013}

\begin{document}

\maketitle

\begin{abstract}
An ab initio calculations have been carried out for examining the curvature effect of small diameter hexagonal boron nanotubes. The considered conformations of boron nanotubes are namely armchair (3,3), zigzag (5,0) and chiral (4,2), and consist of 12, 20, and 56 atoms, respectively. The strain energy is evaluated in order to examine the curvature effect. It is found that the strain energy of hexagonal BNT strongly depends upon the radius, whereas the strain energy of triangular BNTs depends on both radius and chirality.

\keywords cohesive energy,  curvature effect, strain energy, ab-initio calculations, boron nanotube

\pacs 31.15.A-,  06.30.Bp, 61.48.De,  62.25.-g, 71.15.Mb
\end{abstract}

\section{Introduction} \label{Sec:INTRODUCTION}

Boron is an electron deficient element~\cite{pauling196077ie} in the periodic table and possesses a richness of chemistry next to carbon. Boron has a rather fascinating chemistry. Pure boron compounds have neither purely covalent nor  purely metallic character. This results in a chemical versatility, which is unique among the elements of the periodic table.
Various nanostructures of boron are as follows: boron nanotubes (BNTs)~\cite{boustani1997nanotubules,boustani1999NewBoron,PhysRevB.74.035413,PhysRevB.77.041402,singh2008probing,ding2009electronic,kunstmann2005constricted,wang2009new} boron sheet (BS)~\cite{lau2007stability,tang2007novel}, boron clusters~\cite{singh2008probing}, boron nanowires~\cite{otten2002crystalline,wang2003crystalline,cao2002synthesis,cao2001well}, boron nanoribbon~\cite{ding2008electronic} and boron buckyball~\cite{PhysRevLett.98.166804}. The boron sheets of various morphologies viz. hexagonal, triangular, and $\alpha$-boron sheet have been predicted based on ab initio calculations. However, boron sheets have not been yet synthesized experimentally. Further it has been found that $\alpha$-boron sheet is highly stable while hexagonal boron sheet is least stable among these conformations. The boron nanotubes rolled from these BS have also been investigated and consequently it has been predicted that all the BNTs are metallic independent of diameter and chirality. The experimental synthesis of pure BNTs~\cite{ciuparu2004synthesis,liu2010metal} has confirmed the existence of BNTs. Moreover, BNTs have a lot of morphologies such as hexagonal, triangular and $\alpha$-BNTs  in contrast to only one morphology of CNTs i.e., hexagonal.

The Aufbau principle was proposed for elemental boron by Boustani~\cite{PhysRevB.55.16426} which states that stable boron cluster can be constructed of two basic units only: a pentagonal pyramidal $B_{6}$ unit and a hexagonal pyramidal $B_{7}$ unit. Further, it implies that quasiplanar~\cite{boustani1997new},  tubular~\cite{boustani1997nanotubules,gindulyte1998quantum}, convex and spherical~\cite{boustani1997new2} boron clusters are made of these units.  Moreover, the stable BNTs and BS can be constructed of only one unit i.e. hexagonal pyramidal $B_{7}$. The existence of quasiplanar clusters or \emph{sheet} was recently confirmed by experiment~\cite{zhai2003hydrocarbon}, in perfect agreement with earlier theoretical predictions~\cite{boustani1999ab}. Beyond the Aufbau principle, hexagonal structures are also possible. Recently, it has been revealed that the thinnest CNT-like-BNTs are stable as predicted by Dongju~\cite{zhang2006density} which confirmed the existence of hexagonal BNTs. Bezugly et al.~\cite{bezugly2011highly} have predicted the boron nanotube to be highly conductive. The boron nanotubes that we have considered are armchair (3,3), zigzag (5,0), and chiral (4,2) because the CNT of the same chirality has been investigated~\cite{PhysRevB.66.155410,yang2003optical,PhysRevB.66.115416} theoretically as well as experimentally. Therefore, the comparison in the properties of the tubes can be studied in order to distinguish them. It is brought out that zigzag (5,0) CNT is found to be metallic in contrast to semiconducting (as per zone folding prediction). This transition is attributed to the curvature effect. To our best knowledge, curvature effect is also a prominent one at small diameters which can be examined through strain energy calculations. The strain energy is a function of diameter and/or chirality. As the diameter increases, the curvature energy  decreases. Consequently, curvature energy is more dominant at a small diameter. In this paper we have studied the effect of curvature in the hexagonal BNTs of small diameters.

\section{Computational methodology} \label{Sec:COMPUTATIONAL METHODOLOGY }

We have performed pseudopotential plane wave calculations for determining the curvature effect of small diameter BNTs. The geometrical structures of BNTs are optimized and then the total energy of ground state is calculated by using DFT based CASTEP~\cite{clark2005first} simulation tool. The plane wave cutoff energy is set to be 320~eV  and exchange correlation effects are described by a generalized gradient approximation (GGA) proposed by Perdew-Burke-Ernzehrof (PBE)~\cite{perdew1996generalized}. The integration is performed in the first brillouin zone~\cite{monkhorst1976special} by using the k-points generated by $1\times 1\times 8 $ grid parameters. These indicated parameters are sufficient to optimize the structures until the force on each atom becomes less than 0.03~eV/{\AA}. Moreover, self-consistency has been obtained on these parameters for other calculations. The ultrasoft pseudopotential~\cite{vanderbilt1990soft} is used in the reciprocal space. Similar calculations are also separately performed by Mao et al.~\cite{mao2008first} , Wang et al.~\cite{wang2009new}, and Tian et al.~\cite{tian2010ab}.  The intertubular distance is kept fixed 10~{\AA}  in order to avoid a periodic image interaction. The bond length B-B is chosen to be 1.67~{\AA} before optimization. The considered conformations of BNTs include armchair (3,3), zigzag (5,0) , and chiral (4,2) consisting of 12, 20, and 56 atoms, respectively. These tubes are periodic along $z$-axis. The diameters of these tubes are 4.60~{\AA}, 4.78~{\AA}, and 4.87~{\AA} for (5,0), (3,3), and (4,2) BNTs, respectively. In addition, non-spin-polarized calculations are performed.

\section{Results and discussion} \label{Sec:RESULTS AND DISCUSSION}

The ab-initio calculations are carried out for boron nanotubes (BNTs) viz. zigzag (5,0), armchair (3,3), and chiral (4,2) of diameters 4.60~{\AA}, 4.78~{\AA}, and 4.87~{\AA}, respectively, with the number of atoms 20, 12, and 56, respectively. The strain energy of a nanotube is defined as the amount of energy required to bend a sheet into nanotube. Thus, strain energy is the difference of cohesive energies between the sheet and nanotube i.e., $E_\textrm{strain}=E_\textrm{sheet}-E_{NT}(R,\theta)$, where $E_\textrm{sheet}$ and $E_{NT}$ are the cohesive energies~\cite{monkhorst1976special} of the sheet and tube, respectively.
Strain energy $E_\textrm{strain}$ quantifies (1) the difference in cohesive energy among different ($R,\theta$) nanotubes, (2) the deformation energy per atom, which is necessary to roll up a single sheet into a nanotube of certain radius and chiral angle ($\theta$), and (3) it is a measure of the mechanical tension of a nanotube. This tension stabilizes the tubular shape.

For CNTs the strain energy effectively depends on the radius $R$ rather than on the chirality $(\theta)\:$ $E_\textrm{strain}=E_\textrm{strain}(R)$. This radial dependence makes it possible to understand that the radius is just a measure for the curvature of a CNT, and the smaller its radius the more energy is needed to bend a Graphene sheet into CNT. The strain energy is independent of chirality which is attributed to the nearly isotropic in-plane mechanical properties of the Graphene sheet. It was found for CNTs that the diagonal elements of the elastic tensors are the same due to hexagonal symmetry of the honeycomb lattice~\cite{saito1998physical}. Therefore, when stretching a Graphene sheet along different in-plane lattice directions, one observes the same stiffness, and the systems  behave like a homogeneous 2D continuum. From a chemical point of view this mechanical isotropy is caused by a hexagonal network of stiff $sp^{2} \sigma$ bonds. Thus, when rolling up a graphene sheet along different in-plane directions to form various nanotubes with similar radii, this process requires similar deformation energies. Therefore, $E_\textrm{strain}$ will be independent of the chirality of the CNT. This mechanical behavior is analogous to a simple sheet of paper that is rolled up to form a tube. This process will require little energy for big radii and more energy for small radii. However, due to the isotropic in-plane mechanical properties of the paper sheet, the energy needed to roll up a paper tube is independent of the \emph{roll up direction}. A similar behavior is also known for BN~\cite{rubio1994theory,chopra1995boron}, BC$_{3}$~\cite{hernandez1998elastic} and MoS$_{2}$~\cite{seifert2000structure} nanotubes.

Further, it had been reported for triangular boron sheet that stretching the sheet along its armchair direction will be much harder than stretching it along its zigzag direction~\cite{kunstmann2007approach}. The bending of the sheet along the armchair direction, which involves the bending of rather stiff $\sigma$ bonds, will take more energy (strain energy) than bending the BNTs along their zigzag direction, where no $\sigma$  bonds will be affected. It is investigated that the armchair BNTs have high strain energy, whereas zigzag BNTs have nearly vanishing strain energies. Thus, the strain energy of triangular BNTs depends on their radii and on their chiral angles i.e., $E_\textrm{strain} = E_\textrm{strain}(R,\theta)$. Thus, the triangular boron sheet basically behaves like a piece of cloth that is reinforced along one direction with parallel chains of stiffeners (the $\sigma$ bonds). Bending the cloth along the lines of stiffeners (armchair direction) takes much more energy than bending the cloth perpendicular to it (zigzag direction). The calculations further predict that BNTs are always metallic, independent of their radii and chiralities. The Fermi surface of the triangular boron sheet has some well pronounced contours in the 2D Brillouin zone, and backfolding of the Fermi surface into the 1D Brillouin zone of a BNT is possible for any radius and any chirality~\cite{PhysRevB.74.035413}. For Graphene, on the other hand, the Fermi surface just exists at the k-points of the Brillouin zone, and backfolding of these special points into the 1D Brillouin zone of a $(n,m)$ CNT is possible, but only if $(n-m)$ turns out to be a multiple of 3~\cite{saito1998physical}. Thus, for CNTs their electronic properties (semiconducting versus metallic) vary quite strongly with radius and chiral angle, but their energies are independent of chirality. BNTs are just the opposite, in the sense that their electronic properties will not depend on the structure type but their total energies actually do. To our knowledge, this is the first nanotubular system for which the theory predicts a direct control over its basic structural and electronic properties. The spectrum of nanotube radii obtained during the synthesis of CNTs will depend on the specific reaction conditions (temperature, pressure, catalyst, reaction gas, etc.), and it can be shifted and/or broadened by changing these conditions. Nevertheless, the CNT chiralities remain random and rather uncontrollable. This was demonstrated by Iijima et al., Bethune et al. and Journet et al.~\cite{iijima1993single,bethune1993cobalt,journet1997large} , who were all synthesizing single walled CNTs using the arc-discharge method, but they reported different mean diameters of 1.0, 1.2 and 1.4~nm, respectively.

Furthermore, they noted that the chiral angles varied quite strongly for a given tube diameter. This is a direct consequence of the $E = E(R)$ dependence of CNTs, because during synthesis the reaction conditions just determine a certain energy range for the resulting nanotubes which gives a certain range of radii but leaves the chirality totally unspecified. In contrast to this, the energies of nanotubes like BNTs, which are derived from a sheet with anisotropic in-plane mechanical properties, strongly depend on their chiralities and radii $E = E(R,\theta)$, and the reaction conditions will effect both structural parameters. Such a behavior might ultimately allow for better structure control among nanotubular materials, because now the different chiral angles should be energetically separable and thus experimentally accessible.

\begin{figure}[!h]
\centerline{
\includegraphics[width=0.45\textwidth]{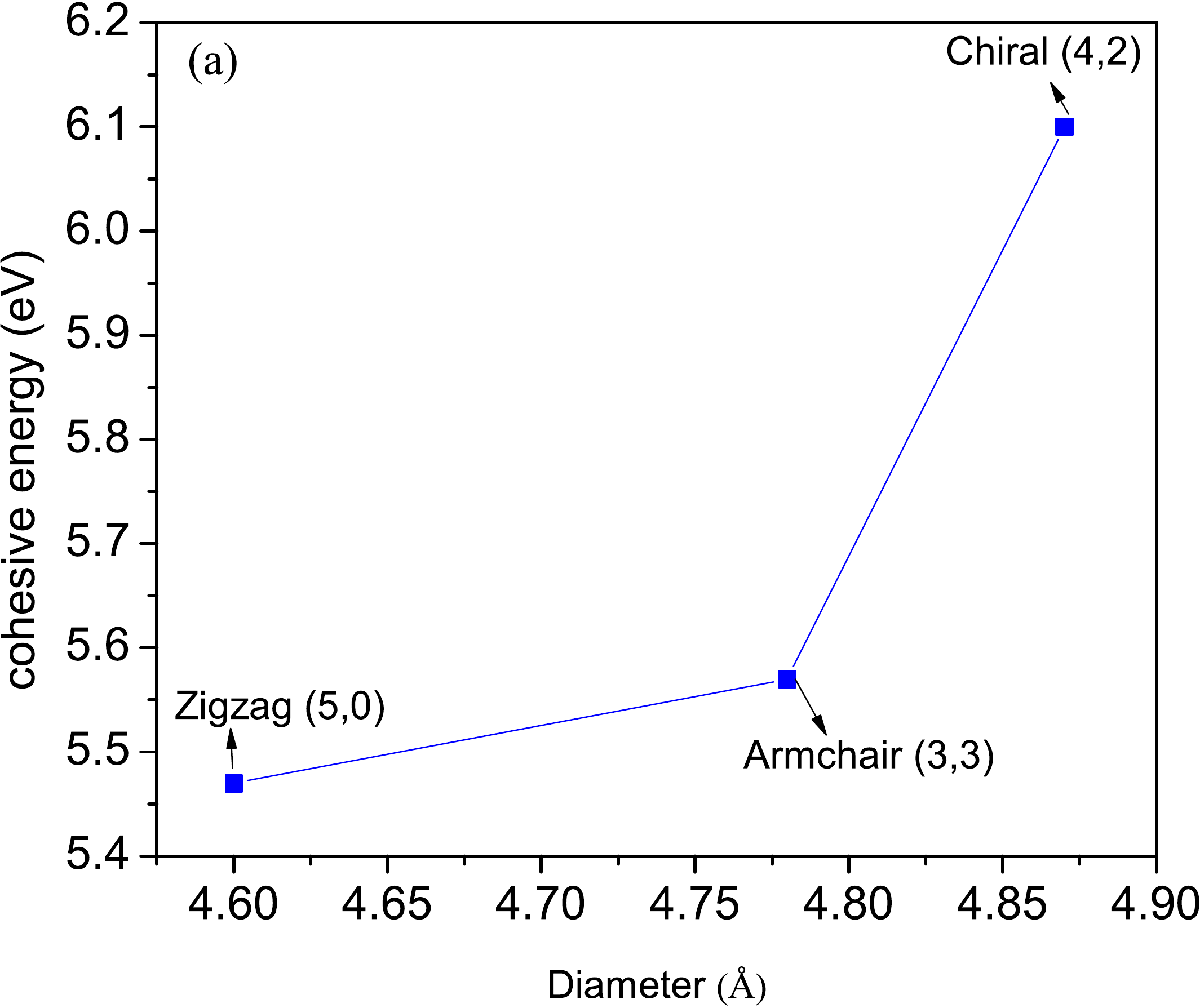}
\hspace{5mm}
\includegraphics[width=0.453\textwidth]{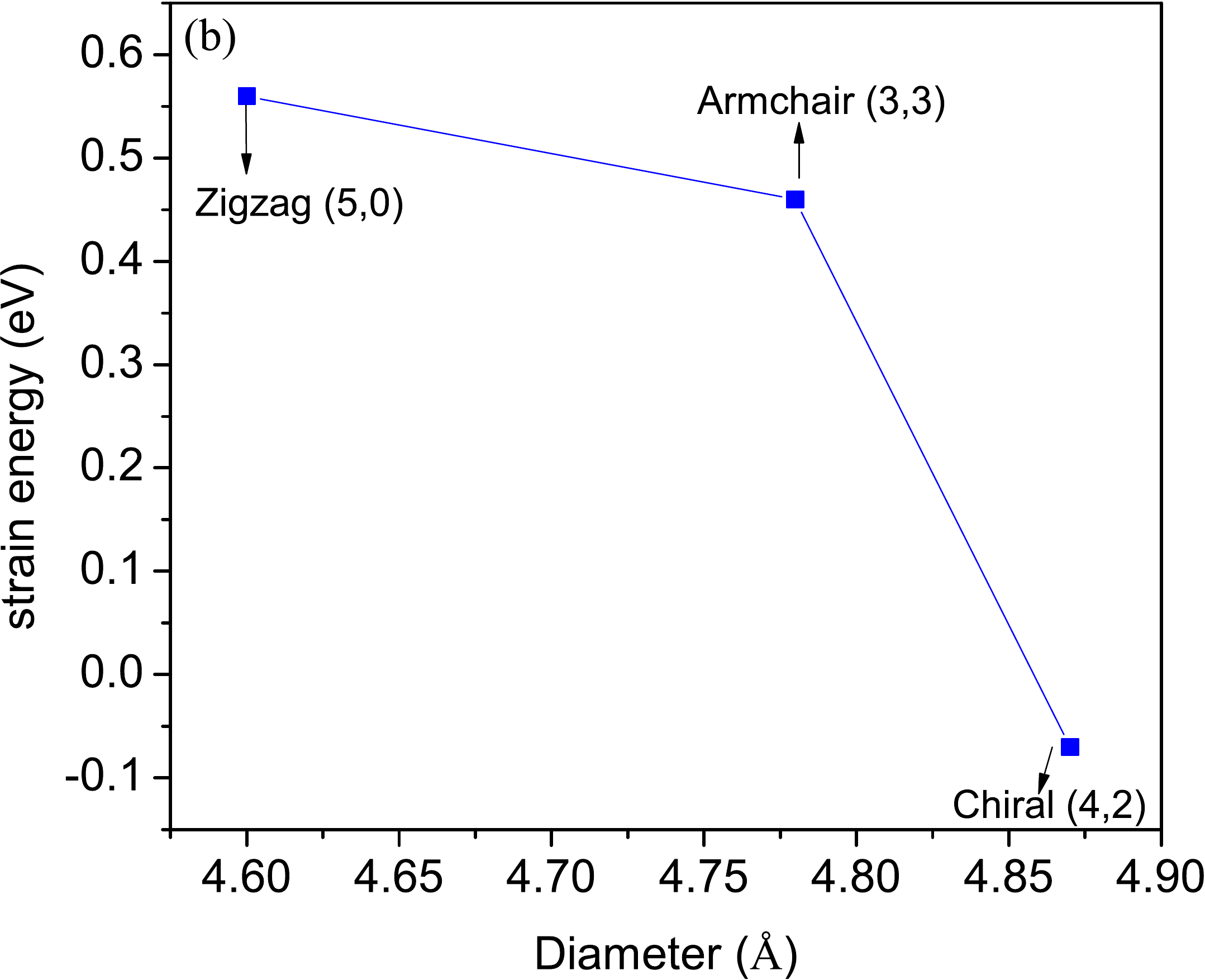}
}
\caption{(Color online) (a) Cohesive energy is shown as a function of diameter, and (b) Strain energy is shown as a function of diameter.}\label{fig:fig(2)}
\end{figure}
In this study, we have considered boron nanotubes (BNTs) of hexagonal lattice structure. Therefore, the chiral angle $\theta$ of BNTs ranges from $0^\circ$ to $30^\circ$. Thus, BNTs and CNTs relate to reference lattices of the same symmetry, and, therefore, one has to use the same chiral indices i.e., ($n, m$) for both CNTs and BNTs. When stretching a boron sheet along different in-plane lattice directions, the same stiffness is observed, and the system behaves like a homogeneous 2D continuum. From a chemical point of view this mechanical isotropy is caused by a hexagonal network of stiff $ sp^2 \sigma $ bonds. Thus, the rolling up of a boron sheet along different in-plane directions to form various nanotubes with similar radii requires similar deformation energies. Therefore, $E_\textrm{strain}$ will be independent of the chirality of the hexagonal BNT. This mechanical behavior is analogous to a simple sheet of paper that is rolled up to form a tube. This process will require little energy for big radii, and it is becoming more and more costly with decreasing radii. But due to the isotropic in-plane mechanical properties of the paper sheet, the energy needed to roll up a paper tube is independent of the \emph{roll up direction}.

It is evident from figure~\ref{fig:fig(2)} that the structural stability depends on diameter. The cohesive energy increases with the diameter. Therefore, it reveals that chiral BNT (4,2) is the most stable and zigzag (5,0) is the least stable while armchair (3,3) is in between the two. The cohesive energies of hexagonal boron sheet and BNTs have been taken (6.03~eV) from references~\cite{bezugly2011highly} and \cite{jain2011electronic}, respectively. It is clearly illustrated that the effect of curvature decreases as the diameter increases and strain energy tends to zero for BNTs of large diameter. Thus, it is evident from figure~\ref{fig:fig(2)} that curvature plays a significant role at small diameter and the zigzag (5,0) BNT possesses high curvature energy followed by armchair (3,3) and chiral~(4,2).

\begin{figure}[!h]
\centerline{
\includegraphics[width=0.45\textwidth]{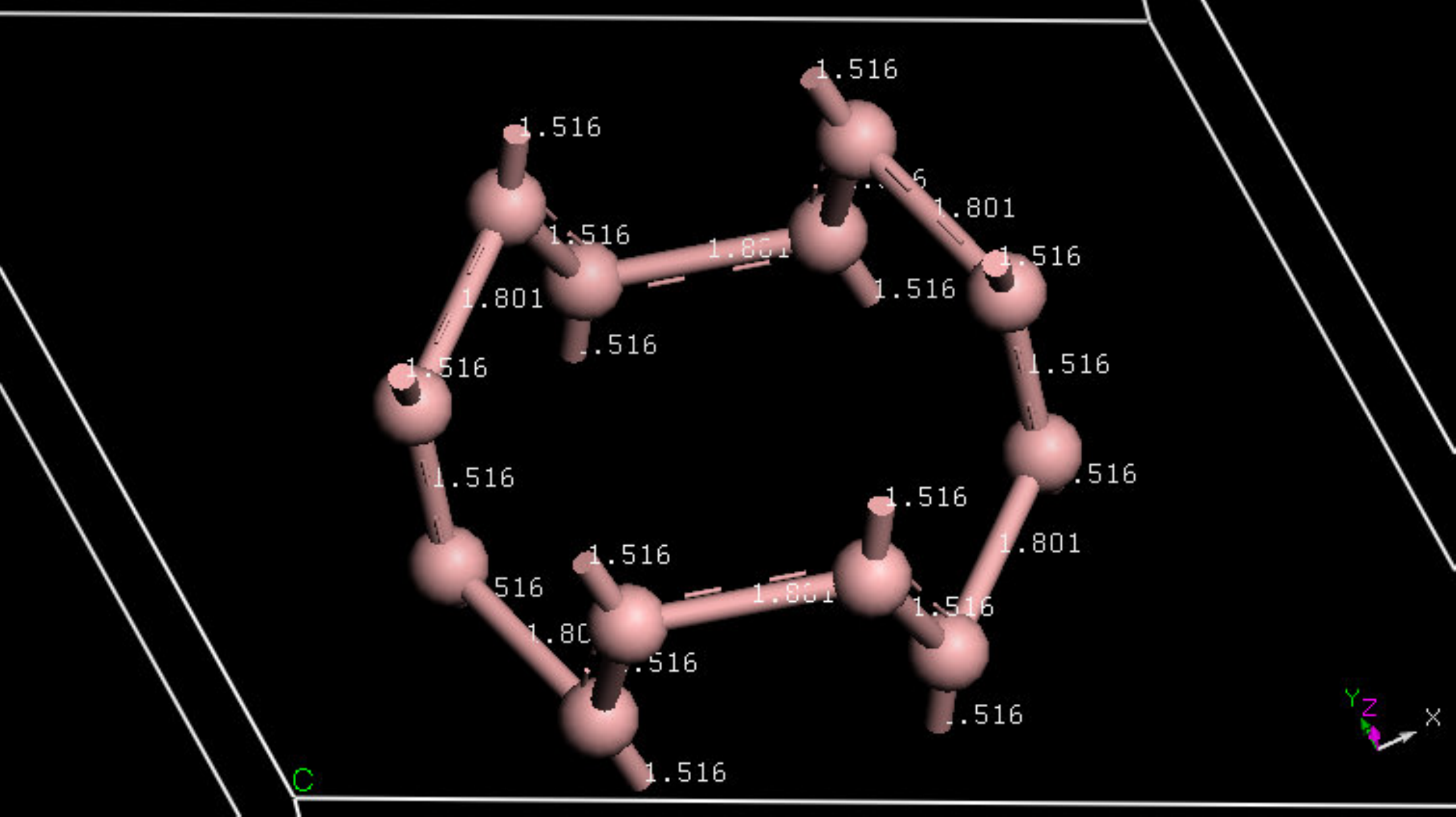}
\hspace{5mm}
\includegraphics[width=0.45\textwidth]{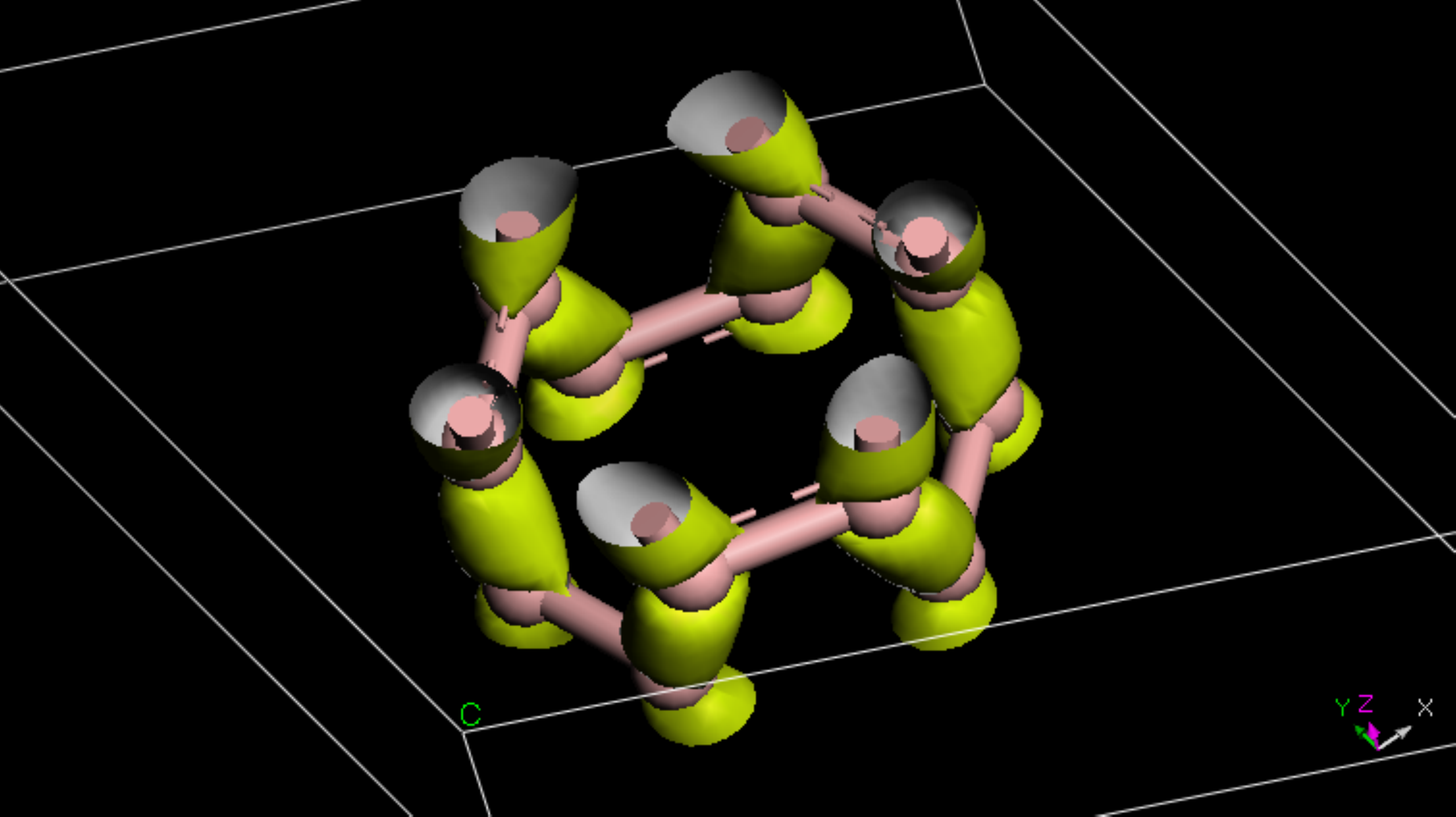}
}
\centerline{(a) \hspace{0.45\textwidth} (b)}
\caption{(Color online) (a) Optimized structure of armchair (3,3) BNT with bond lengths, and (b) Charge density is shown in armchair (3,3) BNT.}\label{fig:fig(3)}
\end{figure}

Furthermore, the charge density distribution of the considered BNTs has been studied in the framework of DFT.  The bond length of optimized armchair (3,3) BNT is explicitly shown in figure~\ref{fig:fig(3)} and brings out that the circular cross section is  extended as the bond length (1.80~{\AA})  is increased while along the axial direction, the bond lengths are contracted to 1.516~{\AA} showing strong bonding. The Charge density distribution of the same BNT shows more density along axial direction  which results in strong bonding.  The charge accumulated between two atoms indicates strong and covalent bonding between them.
\begin{figure}[!h]
\centerline{
\includegraphics[width=0.45\textwidth]{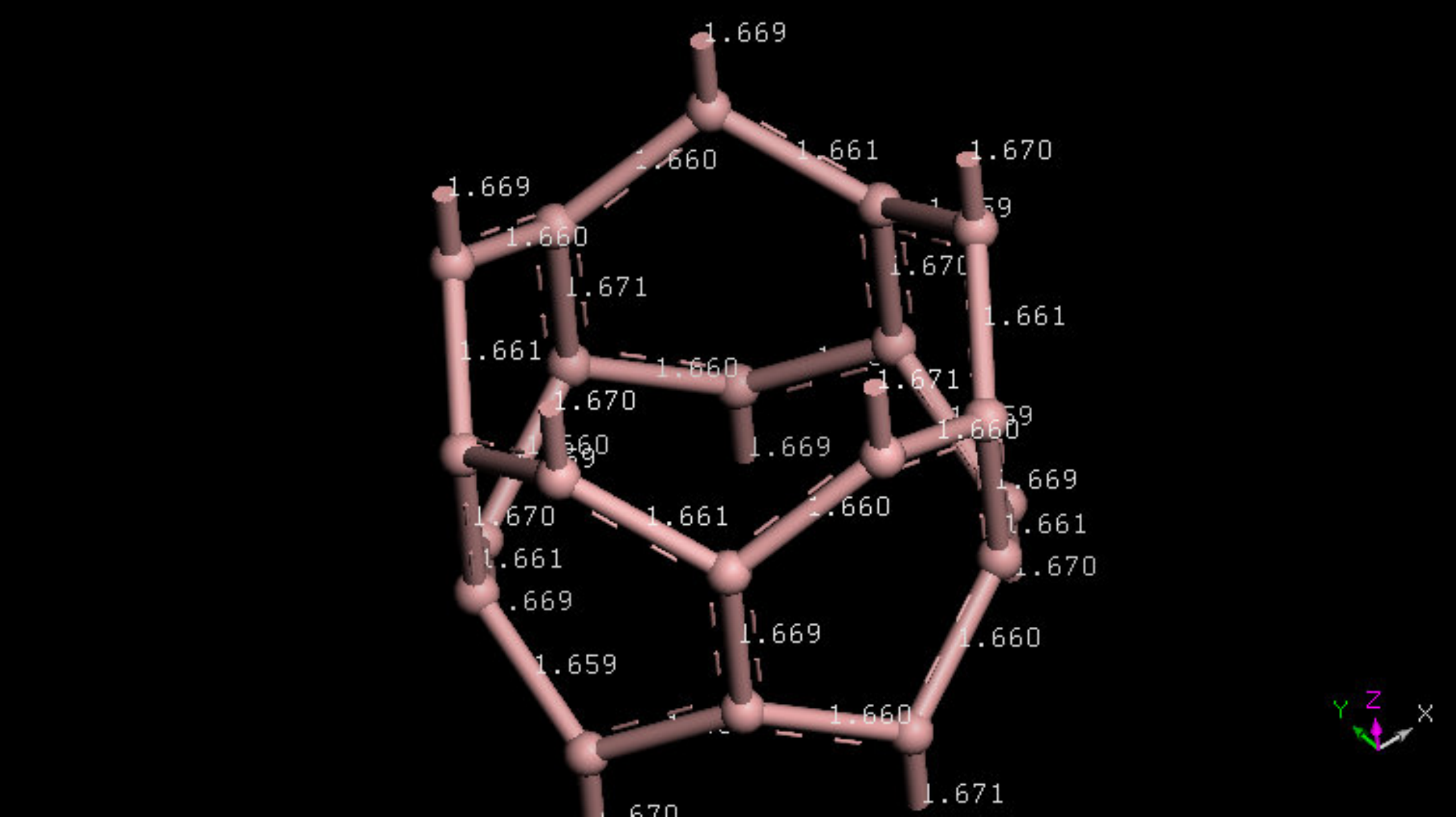}
\hspace{5mm}
\includegraphics[width=0.45\textwidth]{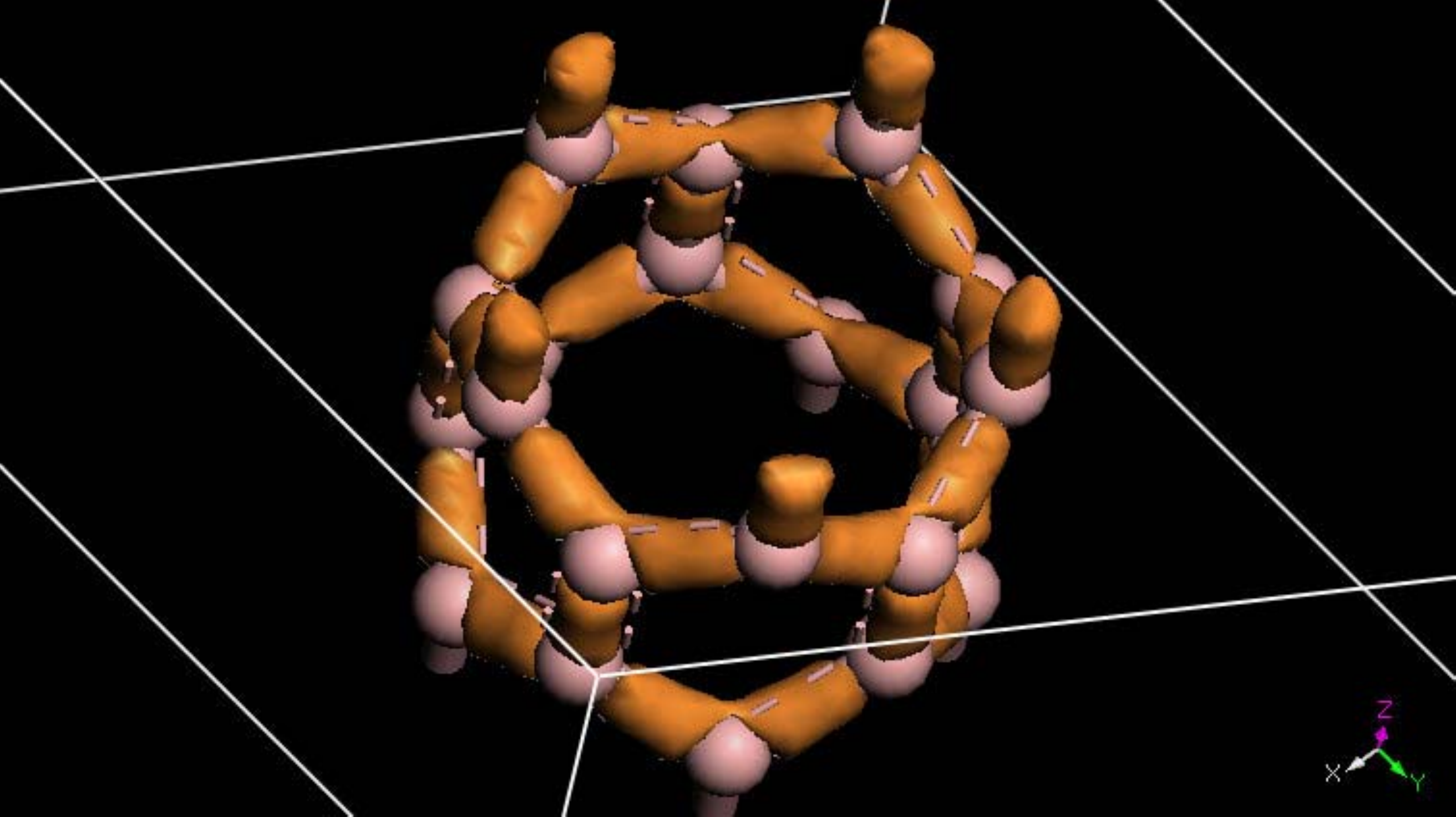}
}
\centerline{(a) \hspace{0.45\textwidth} (b)}
\caption{(Color online) (a) Optimized structure of zigzag (5,0) BNT with bond lengths, and (b) Charge density is shown in zigzag (5,0) BNT.}\label{fig:fig(4)}
\end{figure}

In the case of zigzag BNT, the bond length between atoms is observed to be almost the same i.e., 1.67~{\AA} as depicted in figure~\ref{fig:fig(4)}. The charge density distribution showing the high density between the atoms representing the covalent nature of bonding in zigzag (5,0) BNT. The bond lengths remain the same after optimization.

\begin{figure}[!t]
\centerline{
\includegraphics[width=0.45\textwidth]{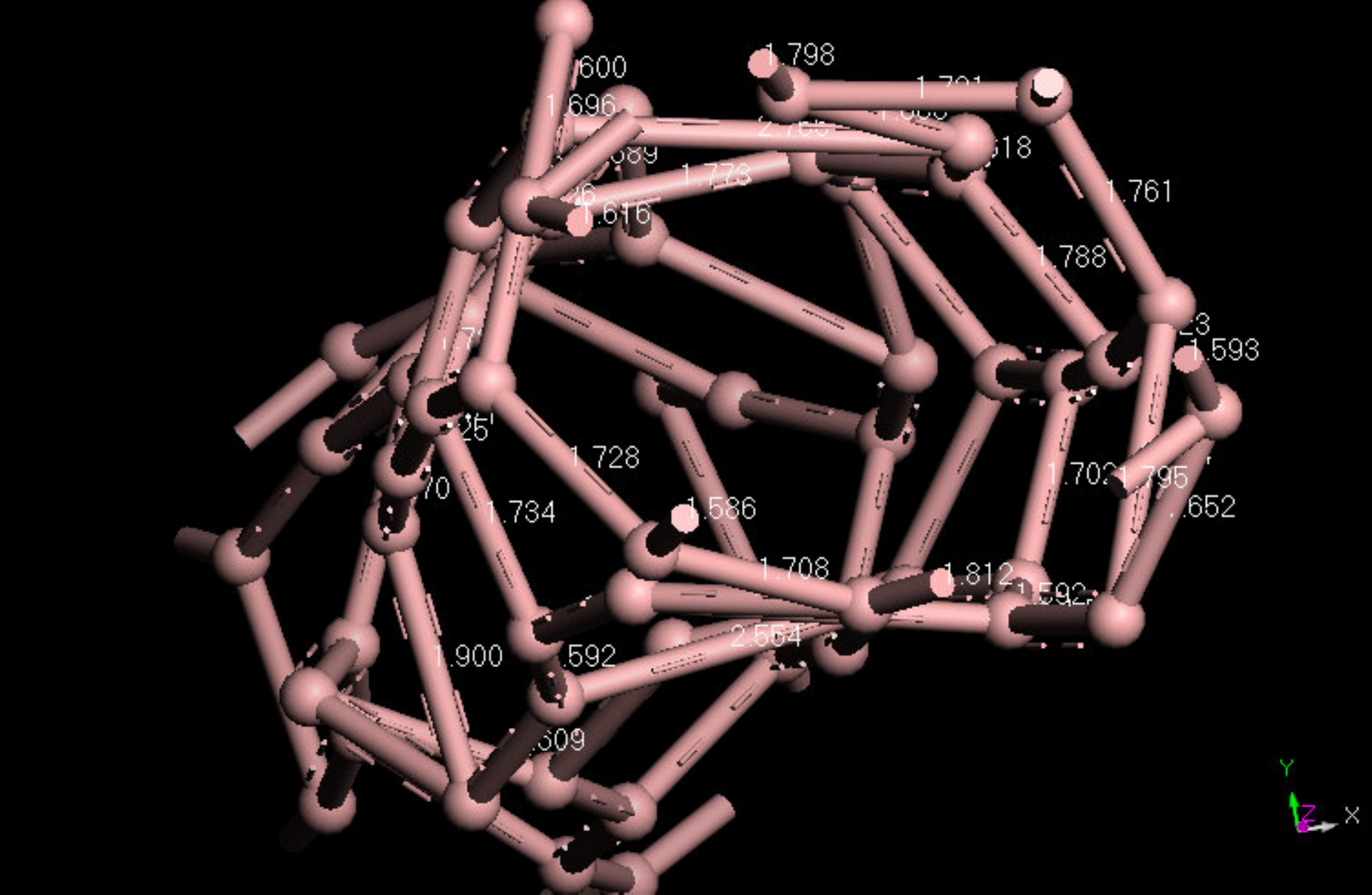}
\hspace{5mm}
\includegraphics[width=0.45\textwidth]{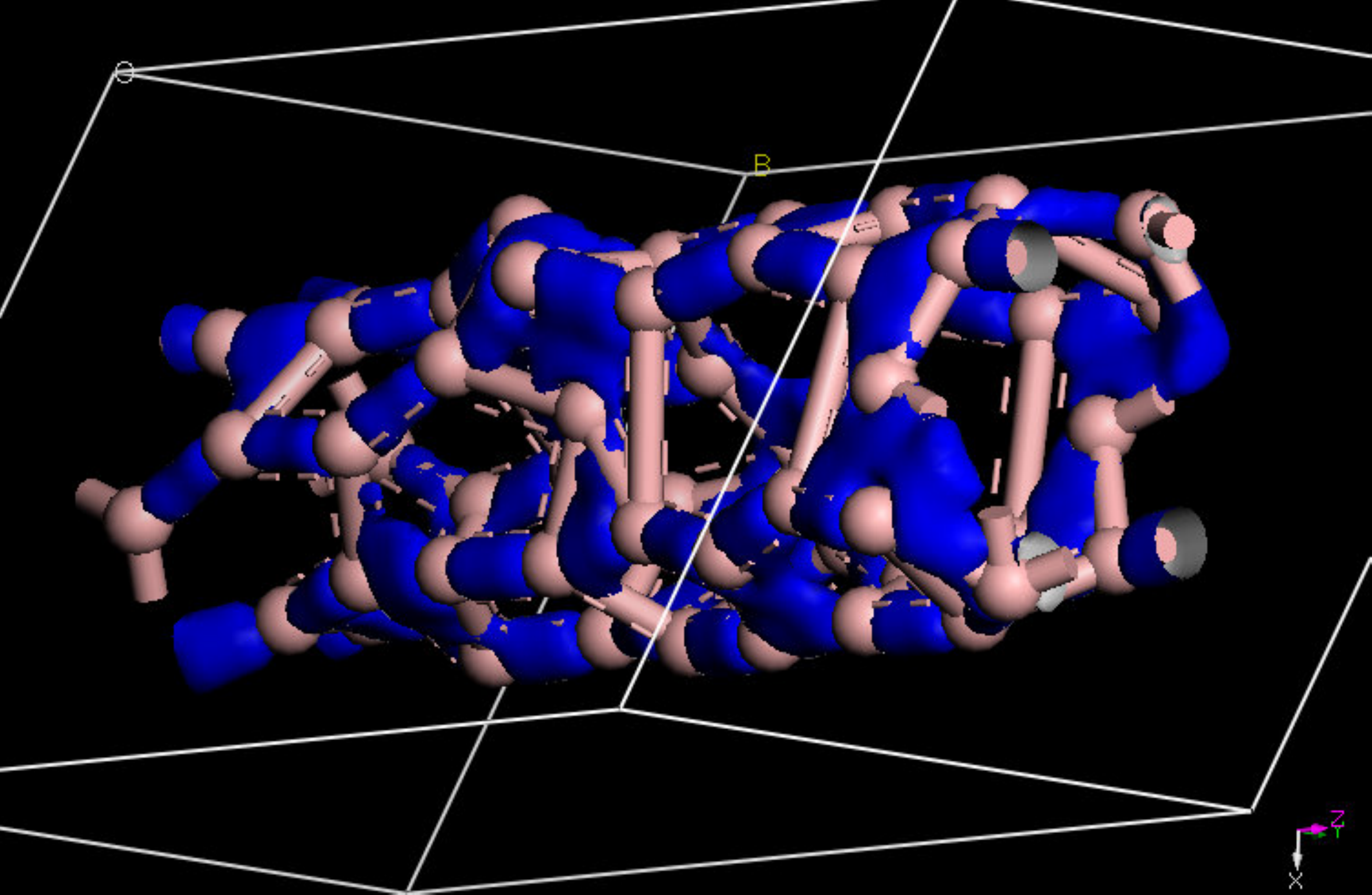}
}
\centerline{(a) \hspace{0.45\textwidth} (b)}
\caption{(Color online)  (a) Optimized structure of chiral (4,2) BNT with bond lengths, and (b) Charge density is shown in chiral (4,2) BNT.}\label{fig:fig(5)}
\end{figure}

The bond lengths along circular cross section are extended which results in an increased diameter and in a curving effect as demonstrated in figure~\ref{fig:fig(5)}. The charge density distribution is also consistent with the optimized structure. The density is greater along the axial direction and results in strong bonding.

\section{Conclusion} \label{Sec:CONCLUSION}

We have carried out ab initio calculations in order to evaluate the strain energy of small diameter boron nanotubes in the framework of density functional theory. We have considered three conformations of BNTs viz. armchair (3,3), zigzag (5,0), and chiral (4,2). All these tubes have nearly the same diameters (i.e., $< 0.5$~nm). Moreover, zigzag (5,0) BNT is the smallest and chiral (4,2) BNT is the largest while armchair (3,3) is in between the two in size. In this paper, we have calculated the strain energy of small BNTs of diameter $< 0.5$~nm and it turns out that the strain energy of chiral BNT is the least followed by armchair and zigzag BNTs. The curvature energy decreases with an increasing diameter. Thus, it is evident that the effect of curvature is noticeable at a small diameter. It is also found that the charge density distribution is also consistent with bond lengths.

\section*{Acknowledgement} \label{Sec:ACKNOWLEDGMENT}
We are very thankful to the Computational Nanoscience \& Technology Lab (CNTL), ABV-Indian Institute of Information Technology \& Management, Gwalior (India) for providing computational facilities. We are also thankful to Accelery Inc for providing the perpetual license of  CASTEP 4.4 simulation tool.

\newpage
\ukrainianpart

\title{Дослідження енергії деформації гексагональних борових нанотрубок: першопринципний підхід}
\author{С.К. Джаін, П. Срівастава}

\address{Група досліджень наноматеріалів, лабораторія обчислювальної нанофізики і технології, \\%
ABV-Індійський інститут інформаційної технології і менеджменту, Гваліяр--474015, Індія}%

\makeukrtitle

\begin{abstract}
\tolerance=3000%
Здійснено першопринципні обчислення для вивчення впливу кривизни
малого діаметра гексагональної борової нанотрубки. Розглянуті
конформації борової нанотрубки є майже кріслоподібними
 (armchair) (3,3), зигзагоподібними (zigzag)
(5,0) і хіральними (4,2), складаючись, відповідно, з 12, 20 та 56
атомів. Для того, щоб дослідити вплив кривизни, оцінено енергію деформації. Знайдено, що енергія деформації гексагональної  борової
нанотрубки строго залежить від радіуса, тоді як енергія деформації
трикутних борових нанотрубок залежить як від радіуса, так і від
хіральності.
\keywords енергія когезії,  вплив кривизни, енергія деформації,
першопринципні обчислення, борова нанотрубка

\end{abstract}

 \end{document}